\documentstyle[11pt,epsfig]{article}
\title{\bf Deflation of Vacuum Energy During Inflation Due to Bulk-Brane Interaction }
\author{ Y. Bisabr \footnote{email: y-bisabr@sru.ac.ir} ~~ and~~ F. Ahmadi\footnote{email:fahmadi@sru.ac.ir} \\{\small Department of Physics, Shahid Rajaee
Teacher Training University, Lavizan, Tehran 16788, Iran.}\\}
\begin{document}
\maketitle
\begin{abstract}
We consider a brane world inflationary model in which inflation is driven by dynamics of a self-interacting scalar field living in the five-dimensional bulk. The scalar field is non-minimally coupled to matter fields on the brane and
acts as an inflaton which induces a slow-roll inflation. We show that although the Friedmann equation is modified at early times due to effects of the extra dimension, the slow-roll condition is the same as that of the four-dimensional case. Due to the non-minimal coupling of matter with the bulk scalar, there is an energy transfer between the two components. We investigate the conditions under which the direction of this energy transfer can be from matter onto the bulk scalar. There are at least two advantages in this case: 1) It establishes a mechanism by which a large effective cosmological term on the brane is deflated during the inflation period. 2) The energy flow onto the bulk inflaton gives a more strongly damped evolution of the scalar field in the slow-roll region for a given potential. We then show that our results are supported by numerical estimations with quadratic and exponential potentials.
\end{abstract}
\vspace{.7cm}
\section{Introduction}
It is widely believed that there are two periods of accelerated expansion phase in evolution of the Universe.
At early times, it is assumed that the Universe has been passed through an
inflationary phase to address some of cosmological puzzles such as flatness
and horizon problems \cite{1-1}. On the other hand, a series of cosmological observations suggest that the Universe has been recently entered an  accelerating expansion phase \cite{1a}. There are various scenarios which have been proposed for explaining this late-time accelerated expansion. The simplest one is appealing to the negative pressure produced by a cosmological constant term in $\Lambda$CDM model \cite{cdm}. One important problem associated with this interpretation is a large contradiction between the late-time observational value of $\Lambda$ and the early-time value obtained by particle physics estimations \cite{1b}. In fact the $\Lambda$-problem is a huge discrepancy between
theory and observations which is an incredible $120$ orders of magnitude \cite{1b} \cite{1} \cite{2}.\\
There have been many attempts trying to resolve the $\Lambda$-problem \cite{1}, which most of them are based
on the belief that $\Lambda$ may not have such an extremely small value suggested by observations at
all the time. In this viewpoint, there should exist a dynamical relaxation mechanism working during evolution of
the universe which has led to reduction of the large $\Lambda$ at early times to a vanishingly small value at late times. Among various possibilities for such a relaxation, one of the most interesting ones is that the energy density associated with $\Lambda$ is attenuated due to an interaction with other energy components in the Universe \cite{3} \cite{4}. The present paper is a short account of such an attenuation in inflationary period. The reason why we consider such a possibility in the period of inflation is that a large $\Lambda$ at early times may affect successful parts of the hot big bang model such as the theories of primordial
Nucleosynthesis or structure formation to be faced with serious difficulties. To avoid such difficulties, the outset of the relaxation mechanism should be pushed to earlier times such as the inflationary era. In this case, one seeks improvements of the standard inflation to solve the $\Lambda$-problem too. \\
We use a brane world scenario in which one assumes that our four-dimensional world is a hypersurface (or brane) embedded in a higher dimensional spacetime (or bulk) with one extra spatial dimension. In this context, the
gravitational field propagates into the bulk while matter systems or Standard Model fields are confined to live on the brane. The basic idea in the brane world models can be extended to scalar-tensor brane models in which the five-dimensional spacetime metric has a scalar field partner for describing gravity in the bulk \cite{ma}. In this case, evolution of the four-dimensional world depends not only on the matter fields localized on the brane but also on the scalar field in the bulk. There are different motivations for introducing a bulk scalar field in brane
world models. This scalar field may be used to formulate a low-energy effective theory \cite{kan} or to
address the gauge hierarchy problem \cite{yang}. Moreover, there is a recent tendency to explore how the bulk scalar affects cosmological perturbations \cite{bucher}, the late-time accelerated expansion \cite{our} and four-dimensional gravitational constant induced on the brane \cite{bru}.
An interesting proposal is the possibility that the inflation of the brane is driven by the scalar field in the bulk \cite{buin}. Inspired by this idea, we consider a model of bulk inflation in which the bulk scalar has a dual role. First, it acts as the inflaton field which generates an accelerated expansion at early times. Second, it can relax a large vacuum energy density produced by Standard Model fields on the brane.\\
The present work is organized as follows: In section 2, we introduce a scalar-tensor brane world
model which its vacuum sector is described by a five-dimensional metric tensor together with a self-interacting scalar field $\phi$. This bulk scalar  controls dynamics of the brane in two ways: via an induced stress-tensor on the brane and a non-minimal coupling to matter. The latter causes a bulk-brane energy exchange.
 We write the field equations for the five-dimensional
metric and then induce them on the brane with appropriate boundary conditions.
 In section 3, we apply the model to early-time evolution of the Universe and show that there is a slow-roll inflation.  We consider the conditions under which a large effective vacuum energy density on the brane can be evacuated. We show that this energy transfer eases the slow-energy condition for a given potential. In section 4, we draw our conclusions.

~~~~~~~~~~~~~~~~~~~~~~~~~~~~~~~~~~~~~~~~~~~~~~~~~~~~~~~~~~~~~~~~~~~~~~~~~~~~~~~~~~~~~~~~~~~~~~~~~~~~~~~~~~~~~~~~~~~~~~~~~~~~~~~~~~~~~~~~~~~~
\section{Basic equations}
 Let us start with the action functional\footnote{We work in the unit system in which $k_5=1$. Also, Latin indices denote 5-dimensional
components $A, B, ...=0, .., 4$ while Greek indices run over
four-dimensional ones $\mu, \nu,... = 0, ..., 3$. The
coordinate $y$ is transverse to the brane.}
\begin{equation}
S=\frac{1}{2}\int d^{5}x \sqrt{-g}\left[R-g^{AB}\nabla_{A}\phi \nabla_{B}\phi -2 V(\phi)\right]+\int d^{4}x
L_{m}(\psi_{m},
 \bar{h}_{\mu\nu})
\label{eq1}
\end{equation}
where $R$ is the Ricci scalar associated with the five-dimensional spacetime metric $g^{AB}$ and $\phi$ is a minimally coupled scalar field with the potential $V(\phi)$. The matter part consists
of some matter fields on the brane, collectively denoted by $\psi_m$, which are coupled with the bulk scalar field via
$\bar{h}_{\mu\nu}=A^{2}(\phi)h_{\mu\nu}$ where $h_{\mu\nu}$ and
$\bar{h}_{\mu\nu}$ are the four-dimensional metrics
on the brane. This coupling is similar to chameleon-matter coupling which gives an effective density-dependent mass to the chameleon scalar field \cite{cham}.  \\
We consider the five-dimensional metric
\begin{equation}
dS^2=g_{AB}dx^{A}dx^{B}=h_{\mu\nu}dx^{\mu}dx^{\nu}+b^{2}(t,y)dy^{2}
\end{equation}
and vary the action (\ref{eq1}) with respect to $g^{AB}$ and $\phi$ which gives
\begin{equation}
G_{AB}=\left(T_{AB}|_{bulk}+T_{AB}|_{brane}\right) \label{eq2}
\end{equation}
\begin{equation}
\Box\phi-\frac{dV}{d\phi}=\beta(\phi)T|_{brane} \label{eq3}
\end{equation}
where
\begin{equation}
T_{AB}|_{bulk}=\nabla_{A}\phi\nabla_{B}\phi-\frac{1}{2}g_{AB}\nabla_{C}\phi\nabla^{C}\phi-g_{AB}V(\phi)
\label{eq4}
\end{equation}
\begin{equation}
T_{AB}|_{brane}=\delta^{\mu}_{A}\delta^{\nu}_{B}
\tau_{\mu\nu}\frac{\delta(y)}{b}\end{equation}
\begin{equation}
\tau_{\mu\nu}=A^{2}(\phi)\tau^{m}_{\mu\nu}
\end{equation}
\begin{equation}
\tau^{m}_{\mu\nu}=\frac{-2}{\sqrt{-\bar{h}}}\frac{\delta
L_{m}}{\delta \bar{h}^{\mu\nu}}
\end{equation}
Here $\beta(\phi)=-\frac{d \ln A(\phi)}{d \phi}$ and $\tau_{\mu\nu}$ is the stress-tensor of matter on the brane. Applying Bianchi identities to (\ref{eq2}) leads to
\begin{equation}
\nabla_{A}T^{AB}|_{brane}=-\nabla_{A}T^{AB}|_{bulk}=-\beta(\phi)T|_{brane}\nabla^{B}\phi
\label{eq6}
\end{equation}
which indicates that neither $T^{AB}|_{brane}$ nor $T^{AB}|_{bulk}$ is conserved due to the interaction between the bulk and the brane. \\
In a cosmological context, we work with the five-dimensional Friedmann-Robertson-Walker metric
\begin{eqnarray}
dS^{2}=
-dt^{2}+\tilde{a}^{2}(t,y)[\frac{dr^{2}}{(1-kr^{2})}+r^{2}(d\theta^{2}+\sin^{2}\theta
d\varphi^{2})]+\tilde{b}^{2}(t,y)dy^{2} \label{eq7}
\end{eqnarray}
where $k=0, +1, -1$. The metric coefficients are subjected to the
conditions
\begin{equation}
\tilde{a}(t,y)|_{brane}=a(t),\hspace{.75cm}\tilde{b}(t,y)|_{brane}=b(t)
\label{eq8}
\end{equation}
with $a(t)$ being the scale factor. To write the bulk field
equations, it is useful to define \cite{bin}
\begin{equation}
F(t,y)\equiv\frac{(\tilde{a}'\tilde{a})^{2}}{\tilde{b}^{2}}-(\dot{\tilde{a}}\tilde{a})^{2}-k\tilde{a}^{2}
\label{eq9}
\end{equation}
where a prime denotes a derivative with respect to $y$. If we take $T^{A}_{B}|_{bulk}=diag[-\rho_{\phi}, P_{\phi}, P_{\phi}, P_{\phi}, P_{T}]$ and $\phi=\phi(t)$, the equation $G_{05}=0$ gives
\begin{equation}
(\frac{\tilde{a}'}{\tilde{a}})(\frac{\dot{\tilde{b}}}{\tilde{b}})-(\frac{\dot{\tilde{a}}'}{a})=0 \label{eq10}
\end{equation}
It has a solution of the form
\begin{equation}
\tilde{b}(t,y)= e^{C_{1}(y)}\tilde{a}'(t,y)
\label{eq11}
\end{equation}
which $C_{1}(y)$ is an arbitrary function of $y$. For (0,0) and (5,5) components of the Einstein equations, we obtain
\begin{equation}
F'=\frac{2\tilde{a}'\tilde{a}^{3}}{3}T^{0}_{0}|_{bulk} \label{eq12}
\end{equation}
\begin{equation}
\dot{F}=\frac{2\dot{\tilde{a}}\tilde{a}^{3}}{3}T^{5}_{5}|_{bulk}
\label{eq13}
\end{equation}
We use the latter two equations and (\ref{eq9}) to write \cite{our}
\begin{equation}
\left(\frac{\dot{\tilde{a}}}{\tilde{a}}\right)^{2}=\frac{1}{6}\rho_{\phi}+\left(\frac{\tilde{a}'}{\tilde{b}\tilde{a}}\right)^{2}
-\frac{k}{\tilde{a}^{2}}+\frac{C_2}{\tilde{a}^{4}}
\label{eq14}
\end{equation}
where $C_2$ is a constant of integration. For writing the gravitational equations on the brane, we define
\begin{equation}
T^{A}_{B}|_{brane}=\frac{\delta(y)}{b}diag[-\rho_{b}, p_{b}, p_{b}, p_{b}, 0]
\label{eq16}
\end{equation}
\begin{equation}
T^{\mu}_{\nu}|_{brane}(x^{\alpha},0)=\lim_{\epsilon\rightarrow 0}
\int^{\frac{\epsilon}{2}}_{-\frac{\epsilon}{2}}T^{\mu}_{\nu}|_{brane}
b~ dy=\tau^{\mu}_{\nu}(x^{\alpha})
\label{eq17}
\end{equation}
and use the junction condition
\begin{equation}
\frac{[\tilde{a}']}{a b}=-\frac{1}{3}\rho_{b}\hspace{.5cm}
\label{eq18}
\end{equation}
Assuming the symmetry $y\leftrightarrow -y$ and taking $[F]=F(0^{+})-F(0^{-})$ as the jump of the function $F$
across $y=0$, the Friedmann equation in the spatially flat case ($k=0$) on the brane takes the form
\begin{equation}
H^{2}=\frac{1}{6}\rho_{\phi}+\frac{1}{36}\rho_{b}^{2}+\frac{C_{2}}{a^{4}}
\label{eq19}
\end{equation}
where $\rho_{\phi}=\frac{1}{2}\dot{\phi}^{2}+V(\phi)$ and $H=\frac{\dot{a}}{a}$ is the Hubble parameter. Comparing (\ref{eq19}) with the  usual Friedmann equation in the four-dimensional cosmology reveals that there is a new term $C_2/a^4$
representing the influence of bulk gravitons on the brane \cite{graviton}, the so-called dark radiation. Moreover, the energy density of matter on the brane $\rho_b$ enters quadratically rather than linearly. The salient feature of (\ref{eq19}) is the contribution of the bulk scalar as $\rho_{\phi}$. This is where scalar-tensor brane theory deviates from the standard picture of the brane cosmology\footnote{In the standard picture, there is no bulk scalar and instead matter energy density appears in the Friedmann equation both linearly and quadratically \cite{brax} \cite{branecho}.}. Here the bulk scalar affects dynamics of the brane via this term and the non-minimal coupling to matter. In particular, the latter causes energy to flow onto or away from the brane. We are interested in this energy exchange in the period of inflation. In that period, the dark radiation term is expected to rapidly
disappear once inflation has commenced so that (\ref{eq19}) is effectively equivalent to
\begin{equation}
H^{2}=\frac{1}{6}\rho_{\phi}+\frac{1}{36}\rho_{b}^{2}
\label{eq19a}
\end{equation}
For writing (\ref{eq3}) and (\ref{eq6}) on the brane, we use (\ref{eq8}) and (\ref{eq17}) and get
\begin{equation}
\ddot{\phi}+3H\dot{\phi}+\frac{\dot{b}}{b}\dot{\phi}+V'=\beta(\phi)(1-3\omega_{b})\rho_{b}
\label{eq20}
\end{equation}
or
\begin{equation}
\dot{\rho}_{\phi}+3H(\omega_{\phi}+1)\rho_{\phi}=Q-\frac{\dot{b}}{b}\dot{\phi}^2
\label{eq21}
\end{equation}
for the bulk scalar and
\begin{equation}
\dot{\rho_{b}}+3H(\omega_{b}+1)\rho_b=-Q-\frac{\dot{b}}{b}\rho_b
\label{eq22}
\end{equation}
for the matter where
$Q\equiv\beta(\phi)(1-3\omega_{b})\dot{\phi}\rho_b$
and $V'\equiv\frac{dV(\phi)}{d\phi}$. The right hand sides of (\ref{eq21}) and (\ref{eq22}) indicate that $\rho_{\phi}$ and $\rho_{b}$ are non-conserved due to the bulk-brane interaction and dynamics of the extra dimension. In the bulk-brane interaction, $\beta(\phi)$ describes intensity of the coupling and is generally an evolving function. Without loss of generality and for seeking simplicity, we will assume that it is a constant parameter. This assumption is equivalent to taking the arbitrary coupling function $A(\phi)$ to have an exponential form $A(\phi)=e^{-\beta\phi}$.
When the extra dimension is allowed to vary with time, one needs an equation characterizing dynamics of $b(t)$. By using (\ref{eq11}) and (\ref{eq18}), this equation can be written as
\begin{equation}
b(t)= b(x^{\alpha},0)=\lim_{\epsilon\rightarrow 0}
\int^{\frac{\epsilon}{2}}_{-\frac{\epsilon}{2}}e^{C_{1}(y)}\tilde{a}' dy=-\frac{C_{3}}{3} a(t) \rho_{b}(t),
\label{eq23}
\end{equation}
where we have taken $e^{C_{1}(\frac{+\epsilon}{2})}=e^{C_{1}(\frac{-\epsilon}{2})}\equiv C_{3}$.

~~~~~~~~~~~~~~~~~~~~~~~~~~~~~~~~~~~~~~~~~~~~~~~~~~~~~~~~~~~~~~~~~~~~~~~~~~~~~~~~~~~~~~~~~~~~~~~~~~~~~~~~~~~~~~~~~~~~~~~~~~~~~~~~~~~
\section{Inflation and vacuum decay }
In brane world scenarios, inflation can be driven by an inflaton living either in the bulk (bulk inflation) or on the brane (brane inflation). Adopting to the first possibility, we investigate general conditions for which the bulk scalar $\phi$ induces inflation on the brane. Combining the Friedman equation (\ref{eq19a}) with (\ref{eq21}) and (\ref{eq22}) gives
\begin{equation}
\dot{H}+H^2=\frac{1}{12}[\frac{Q}{H}-(\rho_{\phi}+3p_{\phi})]+\frac{1}{36}\rho_b
(\rho_b-\frac{Q}{H}-3(\omega_b+1)\rho_b)
\label{eq23a}
\end{equation}
where $p_{\phi}=\frac{1}{2}\dot{\phi}^2-V(\phi)$ and $\dot{b}=0$ is assumed for simplicity. The following cases can be distinguished:\\
\textbf{a) No matter on the brane:} When there is no matter on the brane, (\ref{eq23a}) is reduced to
\begin{equation}
\dot{H}+H^2=-\frac{1}{12}(\rho_{\phi}+ 3p_{\phi})
\label{eq23b}
\end{equation}
This guarantees $\ddot{a}>0$ when
\begin{equation}
p_{\phi}<-\frac{1}{3}\rho_{\phi}
\label{c}\end{equation}
This is the same as violation of the strong energy condition required for the inflaton fluid to produce an inflation in four-dimensional general relativity. It should be emphasized that when inflation is driven by an inflaton on the brane the condition for $\ddot{a}>0$ is $p_{\phi}<-\frac{2}{3}\rho_{\phi}$ \cite{branecho} which is stronger than (\ref{c}).\\
\textbf{b) Matter fields on the brane with minimal coupling ($\beta=0$):} In this case, there are matter fields on the brane which are directly coupled to the metric $h_{\mu\nu}$. This case corresponds to vanishing of the coupling parameter $\beta$. We assume that in the period of inflation the matter sector is dominated by zero point energies produced by all matter fields\footnote{If all energy components on the brane are decomposed as $\rho_b=\rho_{\Lambda}+\rho_{m}$, with $\rho_{\Lambda}$ and $\rho_m$ being vacuum and matter energy densities, then this assumption implies that $\rho_{\Lambda}>>\rho_m$ and thus $\rho_b\approx\rho_{\Lambda}$.}. All contribute to a large effective cosmological constant $\Lambda$ which in a perfect fluid description satisfies an equation of state $p_{\Lambda}=-\rho_{\Lambda}$. Thus equation (\ref{eq23a}) is reduced to
\begin{equation}
\dot{H}+H^2=-\frac{1}{12}(\rho_{\phi}+ 3p_{\phi})+\frac{1}{36}\rho_{\Lambda}^2
\label{eqc1}
\end{equation}
In this case, inflation takes place as a mixture effect of bulk and brane, $\phi$ in the bulk and $\Lambda$ on the brane. This case is more realistic but suffers the usual problem associated with the cosmological term. One should answer the question that how the zero point energies, accumulated in $\Lambda$, is attenuated to a vanishingly small value at late times.\\
\textbf{c) Matter fields on the brane with non-minimal coupling ($\beta\neq0$):} In this case, there are matter fields on the brane which are coupled with the metric $\bar{h}_{\mu\nu}=e^{-2\beta\phi}h_{\mu\nu}$. This corresponds to the case that $\beta\neq 0$. Due to this non-minimal coupling, there is an energy exchange between matter fields and the bulk inflaton. In the following, we study the conditions under which the direction of this energy transfer be from $\Lambda$ to the bulk inflaton. If such a decay process is efficient enough the bulk inflaton would be responsible for both inflation and energy extraction from $\Lambda$. To study such a possibility, we first rewrite
 the equation (\ref{eq22}) (for $\dot{b}=0$ and $\omega_b=-1$)
 \begin{equation}
\dot{\rho}_{\Lambda}=-4\beta\dot{\phi}\rho_{\Lambda}
\label{eq25}
\end{equation}
This can immediately be solved which gives
\begin{equation}
\rho_{\Lambda}=\rho_{\Lambda_{0}} e^{-4\beta\phi}
\label{eq26}
\end{equation}
where $\rho_{\Lambda_0}$ is the large initial vacuum energy density after the Big Bang. On the other hand, the equation (\ref{eq20}) becomes
\begin{equation}
\ddot{\phi}+3H\dot{\phi}+V_{eff}'=0
\label{eq24}
\end{equation}
where $V_{eff}=V+\rho_{\Lambda}$. For $\beta>0$, $\rho_{\Lambda}$ decays and the initial vacuum density $\rho_{\Lambda_0}$ can be reduced to the current observational value $\rho_{\Lambda_{obs}} \sim 10^{-123}\rho_{\Lambda_0}$ so that
\begin{equation}
\rho_{\Lambda} \rightarrow \rho_{\Lambda}(\phi_m)=\rho_{\Lambda_0}e^{-4\beta\phi_m}=\rho_{\Lambda_{obs}}
\label{eq24ba}
\end{equation}
where $\phi_m$ is the minimum of the potential $V_{eff}(\phi)$.
\begin{figure}[ht]
\begin{center}
\includegraphics[width=0.45\linewidth]{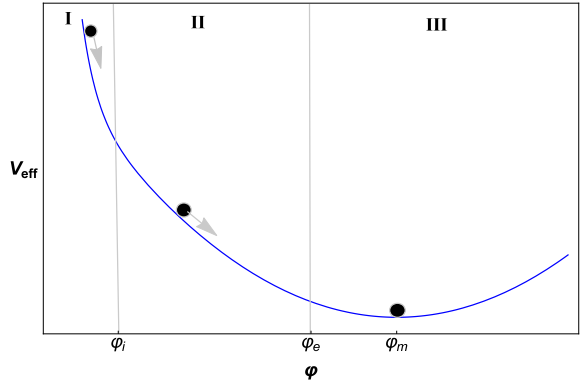}
\caption{A shematic plot of $V_{eff}(\phi)$.}
\end{center}
\end{figure}
Fig. (1) is a schematic illustration of $V_{eff}(\phi)$ and the evolution history during three stages. The stage \textbf{I} is a pre-inflationary era in which the vacuum energy decays into the bulk inflaton. This stage exhibits a relative fast-roll expansion compared to the slow-roll expansion in the next stage. In stage \textbf{II}, the slow-roll inflation is triggered at $\phi_i$ and ends up at $\phi_e$. During this stage, $\rho_{\Lambda}$ continues decaying. The stage \textbf{III} is a reheating phase during which the bulk scalar oscillates around the minimum $\phi_m$ and decays into
standard model particles. A few comments are in order on the slow-roll inflation:\\
Within the slow-roll approximation, it is assumed that the kinetic energy of the inflaton is negligibly small and the energy density $\rho_{\phi}$ is dominated by the self-interacting potential. In four-dimensional General Relativity, this scheme is represented by two equations : one gives the Hubble expansion rate $H^2\propto V(\phi)$ and the other is $3H\dot{\phi}\simeq -V'$ which relates the friction term in the evolution equation of $\phi$ to the slope of the potential. In brane world scenarios in which inflation is driven by an inflaton on the brane, the situation is changed due to contribution of inflaton energy density both linearly and quadratically in the Friedman equation $H^2\propto \rho(1+\frac{\rho}{2\lambda})$ with $\lambda$ being the brane tension. The quadratic term dominates at high energies where the effects of the extra dimension are important while at lower energies it must be sub-dominant in order not to clash with nucleosynthesis \cite{branecho}. The slow-roll expression for the Hubble rate is modified as $H^2\propto V(1+\frac{V}{2\lambda})$ while the evolution equation of the inflaton remains unchanged. It is argued \cite{branecho} that this brane-modification causes the slow-roll inflation to take place more easily for a given potential.\\ Let us look at the situation when the bulk contains a scalar field. In the slow-roll approximation, (\ref{eq19a}) and (\ref{eq24}) take the form
\begin{equation}
H^{2}\simeq\frac{1}{6}V(\phi)+\frac{1}{36}\rho_{\Lambda}^{2}
\label{eq19ab}\end{equation}
\begin{equation}
\dot{\phi}\simeq -\frac{V'_{eff}}{3H}= \frac{1}{3H}(-V'+4\beta\rho_{\Lambda})
\label{eq24b}
\end{equation}
 Since $\rho_{\phi}$ appears linearly in the Friedmann equation, contribution of $\phi$ in (\ref{eq19ab}) is similar to the four-dimensional case. The difference is the second term on the right hand side which gives a faster exponential expansion rate.
A more interesting feature of the above slow-roll equations is related to (\ref{eq24b}). It indicates that the rolling-velocity $\dot{\phi}$ is related to the slope of $V_{eff}$ (rather than $V$) due to non-minimal coupling of matter on the brane. The equation reduces to the usual relationship between $\dot{\phi}$ and $V'$ only for $\beta=0$. For $\beta\neq 0$, the rolling-velocity depends not only on the shape of the potential $V$ but on how $\phi$ interacts with matter on the brane. When $\beta>0$, $\dot{\phi}$ is smaller than the rolling-velocity in the cases $\beta<0$ and $\beta=0$. Since $\dot{\phi}$ governs the size of the kinetic term in $\rho_{\phi}$, one deduces that \emph{the energy flow onto the bulk inflaton assists the slow-roll condition to be satisfied more easily for a given potential}. This behavior is numerically illustrated in fig.2.
\begin{figure}[ht]
\begin{center}
\includegraphics[width=0.45\linewidth]{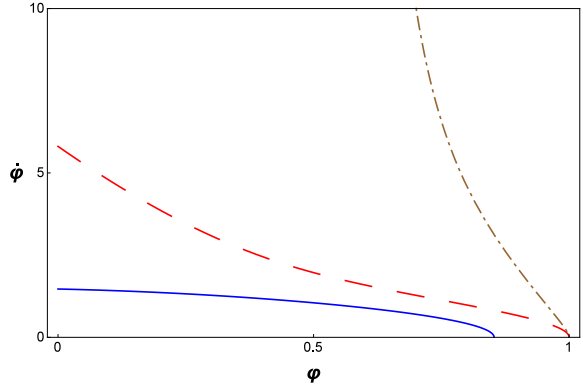}
\caption{A numerical evaluation of $\dot{\phi}$ for a quadratic potential and for $\beta=1/2$ (solid), $\beta=0$ (dashed) and $\beta=-1/2$ (dashdotted).}
\end{center}
\end{figure}
Now we apply the model to quadratic $V_1(\phi)=V_1 \phi^2$ and exponential $V_2(\phi)=V_2 e^{\gamma\phi}$ potentials where $V_1$ and $V_2$ are positive constants and $\gamma$ is a parameter. It should be noted that although the term $e^{-4\beta\phi}$ in the effective potential is monotonically decreasing, by choosing an appropriate $V(\phi)$, the resulting $V_{eff}$ does exhibit a local minimum required for the reheating period. Moreover, it is evident from the definition of $V_{eff}$ and (\ref{eq20}) that both the minimum and its corresponding mass depend on  local matter density. This behavior allows the coupling parameter $\beta$ be of order unity while all local gravitational experiments are satisfied \cite{cham}.\\
\begin{figure}[h]
\begin{center}
\includegraphics[width=0.45\linewidth]{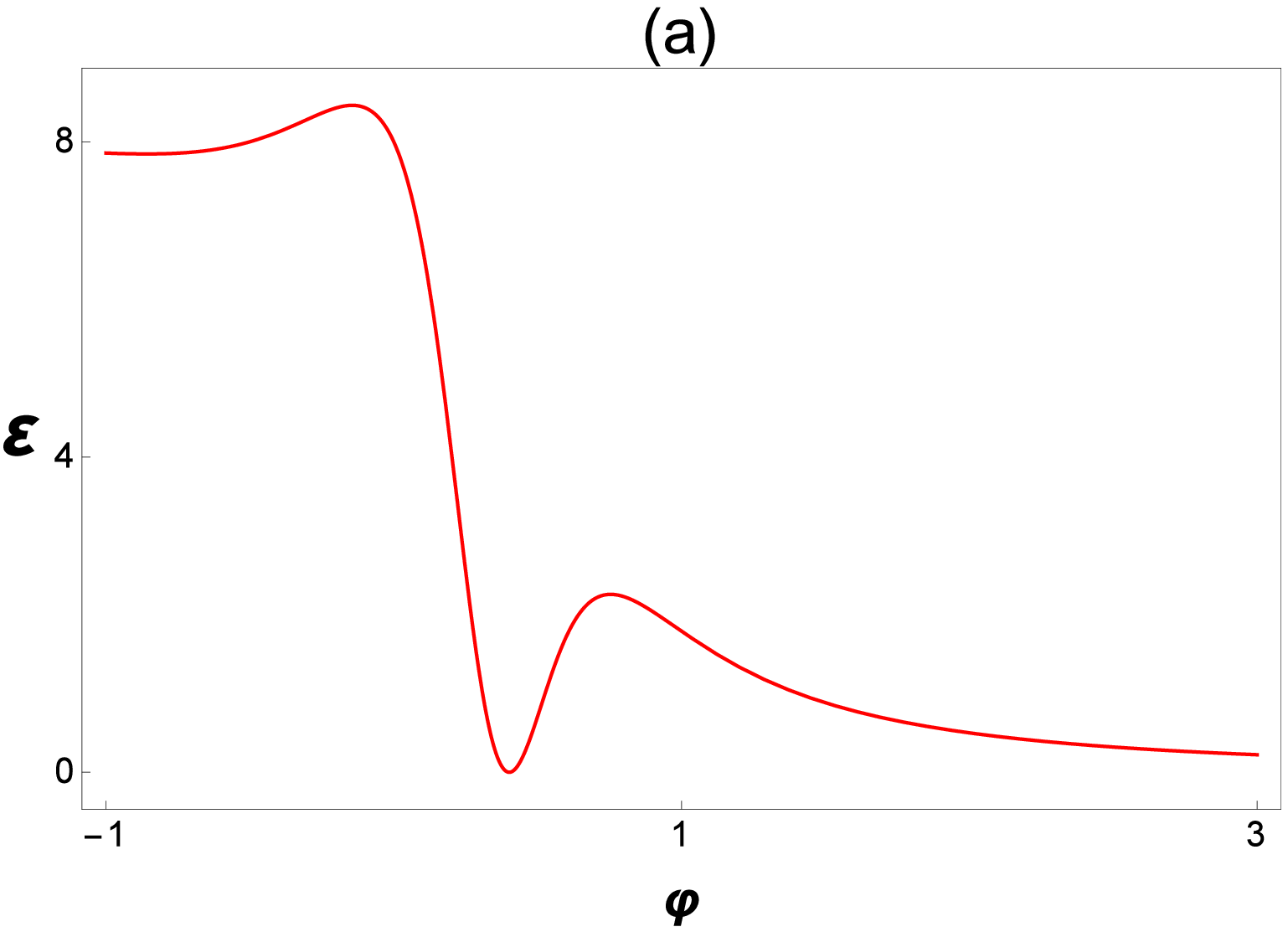}
\includegraphics[width=0.45\linewidth]{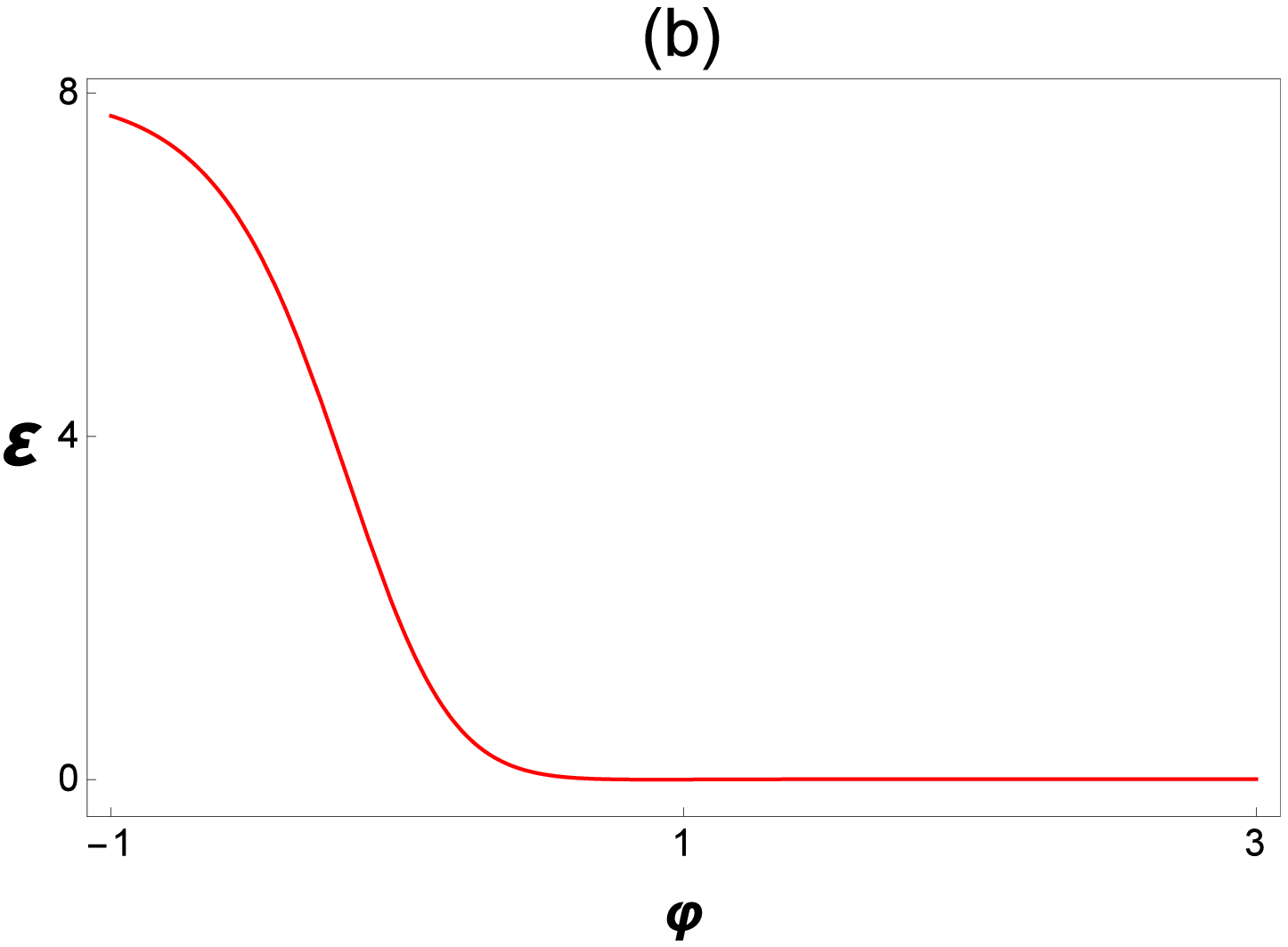}
\caption{Evolution of $\varepsilon(\phi)$ for $\beta=1$ and for $V_1(\phi)$ (pannel a) and $V_2(\phi)$ with $\gamma=0.1$ (pannel b). It is assumed that $V_1=V_2=\rho_{\Lambda_0}$. }
\end{center}
\end{figure}
\begin{figure}[h]
\begin{center}
\includegraphics[width=0.45\linewidth]{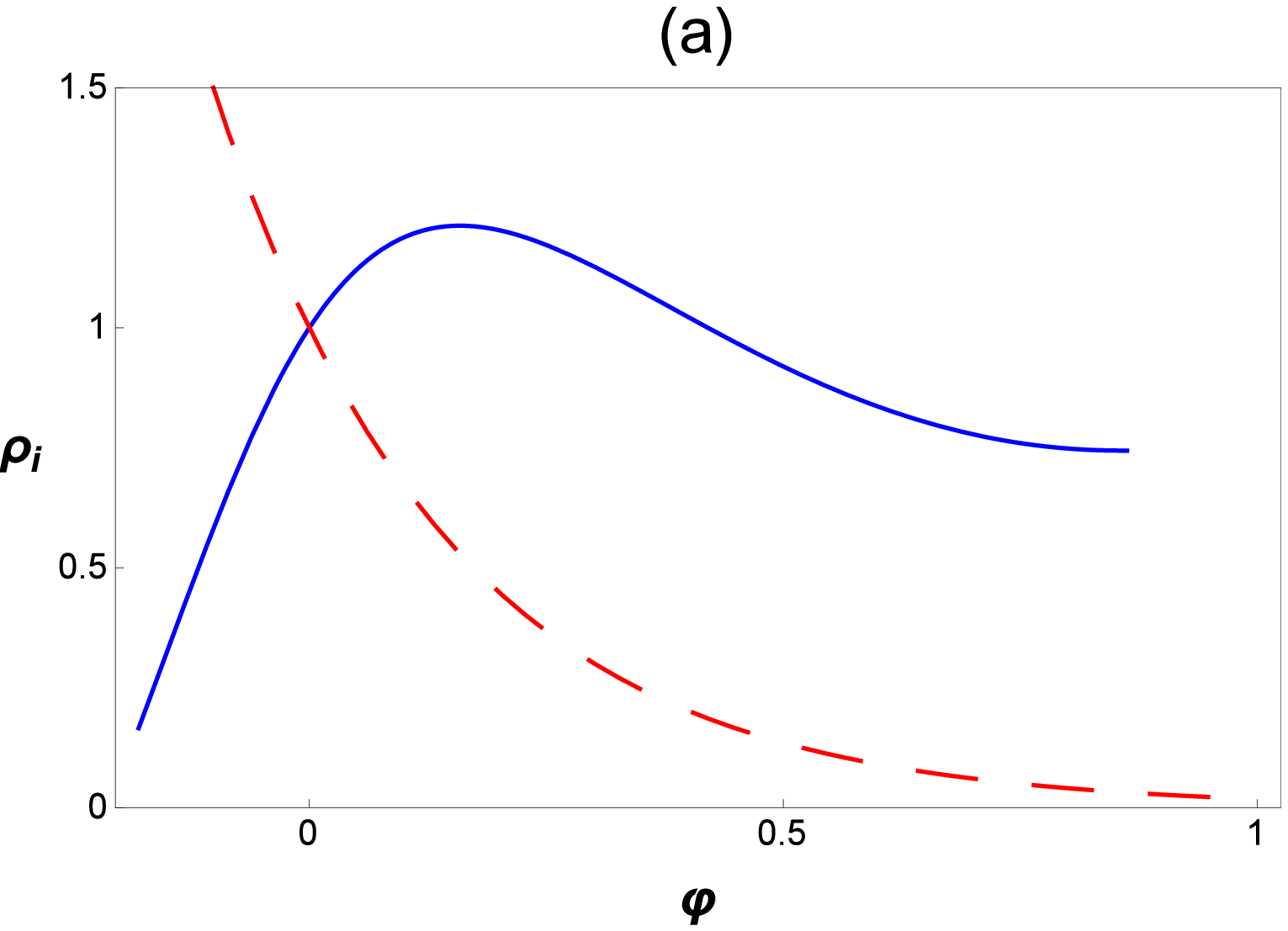}
\includegraphics[width=0.45\linewidth]{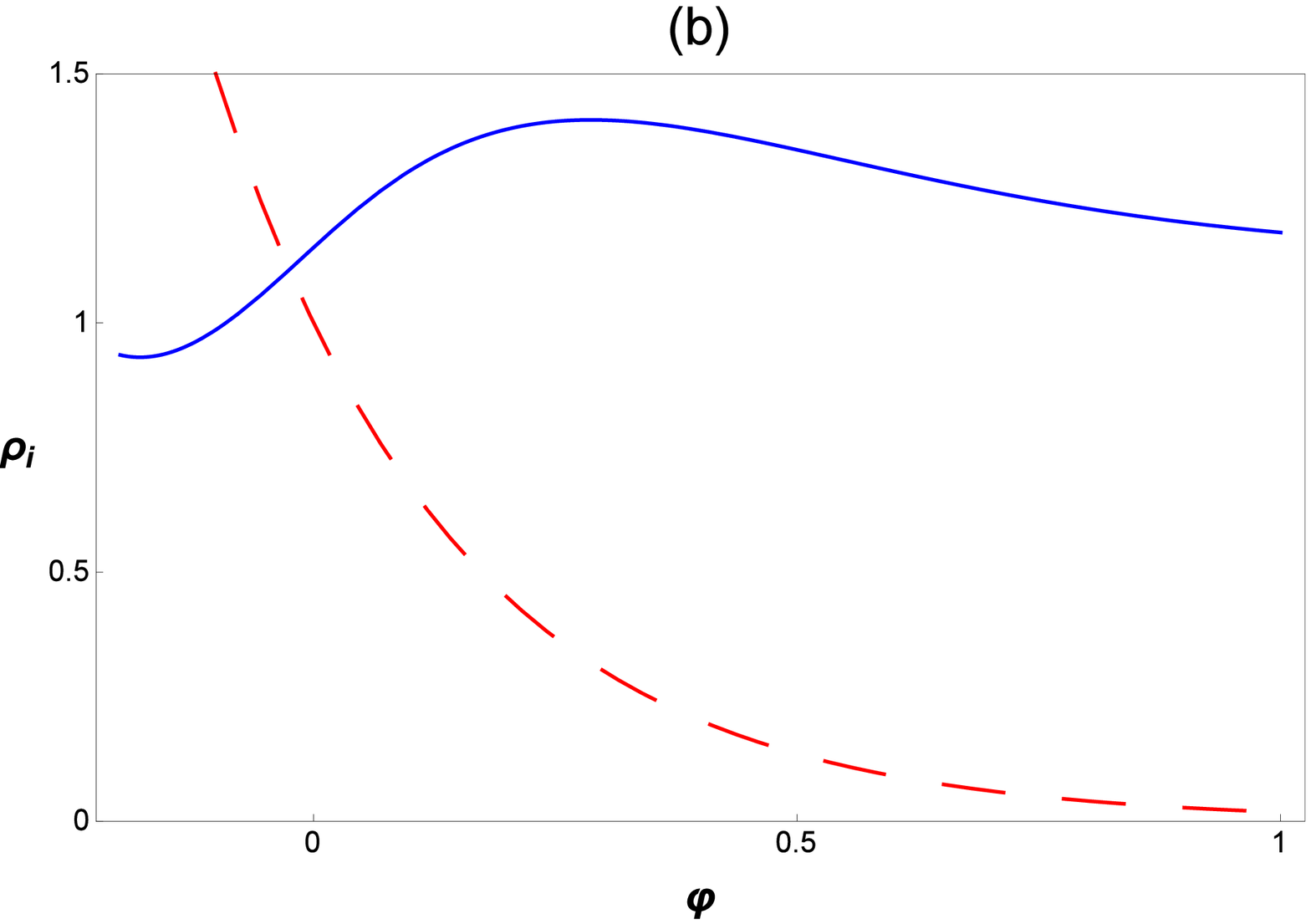}
\caption{Evolution of $\rho_i(\phi)$ for $\beta=1$ with $i=\phi$ (solid) and $i=\Lambda$ (dashed). In these plots we have used $V_1(\phi)$ in pannel (a) and $V_2(\phi)$ with $\gamma=0.1$ in pannel (b). Each plot is scaled such that $V_1=V_2=\rho_{\Lambda_0}=1$. }
\end{center}
\end{figure}
In fig.3, the first slow-roll parameter $\varepsilon=\frac{1}{2}(\frac{V'_{eff}}{V_{eff}})^2$ is plotted
for $V_1(\phi)$ and $V_2(\phi)$. The plots indicate a fast-roll to slow-roll smooth transition. To present the behavior of $\rho_{\Lambda}$ and $\rho_{\phi}$ in pre-inflation and inflation eras, we illustrate $\rho_{\Lambda}$ and a numerical evaluation of $\rho_{\phi}$ in fig.4.
It is evident from the figure that $\phi$ heats up due to decaying of the vacuum energy in the pre-inflationary era. During inflation, $\rho_{\Lambda}$ continues decaying while $\rho_{\phi}$ goes to a constant configuration. This behavior can be realized by exploring the evolution equation of $\rho_{\phi}$, namely
\begin{equation}
\dot{\rho}_{\phi}+3H(\omega_{\phi}+1)\rho_{\phi}=4\beta \dot{\phi}\rho_{\Lambda}
\label{last}
\end{equation}
When $\rho_{\Lambda}\rightarrow 0$ and $\omega_{\phi}\approx -1$ (demanded by the slow-roll condition), (\ref{last}) indicates that $\rho_{\phi}$ goes to a constant configuration.

\section{Conclusion}
There is a class of brane world models which entails a self-interacting scalar field as a partner of the five-dimensional metric in the bulk. This scalar field affects dynamics of the brane via the modified Friedmann equation in which energy densities of the bulk scalar and matter fields respectively appear linearly and quadratically. In this class of brane world scenarios, we have considered a non-minimal coupling between matter fields on the brane and the bulk scalar. Due to this non-minimal coupling, dynamics of the bulk scalar is controlled by an effective potential which its corresponding mass depends on local
matter density. Thus the coupling parameter $\beta$ need not be small and values of order unity or greater
are allowed. \\
We applied the field equations of the model to early-time evolution of the Universe. In particular, we have investigated inflation of the brane driven by the bulk scalar field. This bulk inflation differs from previous studies in this context since dynamics of the bulk inflaton in our case is affected by $V_{eff}(\phi)$ rather than $V(\phi)$, the self-interacting potential. We have shown that evolution of $\phi$ can be divided into three stages: fast-roll (fast compared to the next slow-roll phase), slow-roll and reheating. It is important that existence of a fast-roll period with a smooth transition to a slow-roll phase has been recently proposed \cite{recent} to produce a higher tensor
mode perturbations in order to resolve some inconsistencies between PLANCK and BICEP2 data \cite{recent2}.\\ In our analysis, there is an energy flow onto ($\beta<0$)/ away ($\beta>0$) from the brane during these three stages. The case $\beta>0$ has two interesting features: 1) In this case, a large effective vacuum energy density on the brane can be evacuated into the bulk scalar. Due to exponential decay of $\rho_{\Lambda}$, most of this vacuum evacuation is done in the fast-roll phase. The process continues to happen in the slow-roll and the reheating phases until $\phi$ is settled down in the minimum of the effective potential. 2) Another interesting feature of the case $\beta>0$ is that it gives the smallest possible rolling velocity $\dot{\phi}$ with respect to the other cases $\beta<0$ and $\beta=0$. Thus the energy flow onto the bulk scalar assists the slow-roll inflation for a given potential. Our results have been supported by the
numerical analysis for the two popular self-interacting potentials $V_1(\phi)=V_1\phi^2$ and $V_2(\phi)=V_2e^{\gamma\phi}$.\\\\\\
{\bf Acknowledgment}\\\\
This work is supported by Shahid Rajaee Teacher Training University with a Grant No.29328.


\end{document}